**Entropy Engineering of BF-BT-Based High-Entropy Ceramics for Ultra-High Energy Storage Performance**

Yitao Jiao[1*], Zhenhao Fan[2], Hai-Feng Li[1], Dawei Wang[2]

1. Institute of Applied Physics and Materials Engineering, University of Macau, Avenida da Universidade, Taipa, Macao Special Administrative Region of the People's Republic of China 999078, P.R. China

2. Precision Acousto-optic Instrument Institute, School of Instrumentation Science and Engineering, Harbin Institute of Technology, Harbin 150080, P.R. China

[*]Corresponding author.

E-mail address: yc27825@um.edu.mo

## Abstract

Dielectric capacitors are critical components in pulsed power and advanced electronic systems; however, most lead-free bulk ceramic dielectrics suffer from an intrinsically low recoverable energy density ($W_{rec}$) due to limited breakdown strength (BDS) and pronounced ferroelectric hysteresis. To overcome these limitations, an entropy engineering strategy is proposed by introducing a high-entropy perovskite oxide, $Ba(Ti_{0.2}Zr_{0.2}Sn_{0.2}Hf_{0.2}Nb_{0.1}Sc_{0.1})O_3$ (BTHE), into the $0.7BiFeO_3$–$0.3BaTiO_3$ (BF–BT) ferroelectric matrix, forming a series of BF–BT–xBTHE lead-free high-entropy ferroelectric ceramics. Structural analyses reveal that multicomponent substitution at the B-site induces severe lattice distortion and enhanced pseudo-cubic phase characteristics, leading to pronounced relaxor ferroelectric behavior and significant grain refinement. As a result, both polarization hysteresis and electrical conduction are effectively suppressed, leading to a remarkable enhancement in BDS. An ultrahigh BDS of 840 kV $cm^{-1}$ is achieved, giving rise to a superior $W_{rec}$ of 10.55 J $cm^{-3}$ in bulk ceramics. Furthermore, statistical analysis and phase-field simulations indicate that entropy-induced microstructural heterogeneity effectively disperses local electric fields and delays dielectric breakdown. This work demonstrates that entropy engineering offers a powerful and generalizable approach to simultaneously enhancing BDS and energy-storage performance in lead-free ferroelectric ceramics.

## 1. Introduction

Dielectric capacitors play a pivotal role in pulsed power systems, electric vehicles, renewable energy conversion, and advanced electronics owing to their rapid charge–discharge capability and high power density [1]. The continuous miniaturization and performance advancement of modern electronic systems demand dielectric materials. These materials must simultaneously exhibit high recoverable energy storage density ($W_{rec}$), high energy efficiency ($\eta$), and robust reliability under high electric fields [2]. Although polymer-based dielectrics have demonstrated remarkable energy storage performance exceeding 10 J/cm$^3$, their limited thermal stability and relatively low operating temperature restrict their application in harsh environments [3][4][5]. In contrast, ceramic dielectrics offer superior thermal robustness [6] and long-term reliability, yet their energy storage density remains significantly lower, particularly for lead-free bulk ceramics [7].

Among various lead-free ferroelectric systems, $BiFeO_3$–$BaTiO_3$ (BF–BT) solid solutions have attracted considerable attention due to their high Curie temperature [8], large spontaneous polarization [9], and environmentally benign composition [10]. Previous studies have demonstrated that compositional tuning near morphotropic or polymorphic phase boundaries can enhance electromechanical and dielectric properties in BF–BT ceramics [11]. However, the strong ferroelectric nature of BF–BT typically leads to large remanent polarization ($P_r$) and wide hysteresis loops, resulting in low $\eta$ and limited $W_{rec}$. Recent studies have demonstrated that incorporating linear dielectric components into ferroelectric matrices, such as those in the $BaTiO_3$–$(Na_{0.5}Bi_{0.5})TiO_3$ systems, effectively suppresses hysteresis and improves energy storage performance. This underscores the significance of polarization modulation strategies in dielectric ceramics [12]. More critically, the relatively low BDS of bulk BF–BT ceramics remains a critical bottleneck that is difficult to overcome through conventional doping strategies. [13]. Recently, high-entropy materials, characterized by multiple principal elements occupying equivalent crystallographic sites, have emerged as a promising platform for tailoring functional properties through entropy-driven stabilization and lattice distortion effects [14]. In addition, recent advances in machine learning-assisted materials design have also facilitated efficient phase prediction and performance optimization in high-entropy systems. These developments offer new insights into the complex composition–structure relationships in multicomponent ceramics [15][16]. In ceramic dielectrics, high configurational entropy has been shown to induce severe lattice distortion, sluggish diffusion, and enhanced defect tolerance, which collectively contribute to grain refinement, suppression of electrical conduction, and improved BDS. From a thermodynamic perspective, the configurational entropy ($\Delta S_{config}$) of a multicomponent solid solution can be expressed as [17][18][19]:

$$\Delta S_{config} = -R \sum_{i=1}^{n} x_i ln x_i \quad (1)$$

where $R$ is the gas constant, $x_i$ is the mole fraction of the i-th constituent occupying an equivalent crystallographic site, and n represents the total number of constituent elements. When multiple cations with comparable concentrations are incorporated into a single lattice site, the configurational entropy is significantly increased, stabilizing the solid solution through entropy-driven stabilization [20][21]. Such high-entropy states are thermodynamically favorable at elevated temperatures and can effectively suppress phase separation and long-range ordering [22][23]. In perovskite oxides, the resulting chemical disorder and lattice distortion generate random local fields and spatially fluctuating potential landscapes, which are widely recognized as key factors for inducing relaxor ferroelectric behavior and enhancing dielectric BDS [24][25].

High-entropy concepts have been extensively studied in $BaTiO_3$-based systems and multilayer capacitors; however, their application in bulk BF–BT ceramics remains limited. This is particularly evident in the challenge of achieving both high breakdown strength and high energy storage density simultaneously. In comparison to conventional systems, BF–BT ceramics inherently exhibit strong ferroelectricity and high leakage current, presenting a challenging yet valuable platform for entropy engineering [26].

Recent advances have demonstrated that high-entropy engineering provides an effective pathway to improving the energy storage performance of lead-free dielectric ceramics by simultaneously regulating polarization behavior and BDS. For instance, Liu *et al*. reported a high-entropy $BaTiO_3$-based superparaelectric relaxor ferroelectric system featuring an ultralow $P_r$ (~ 0.58 mC cm$^{-2}$) and a high BDS of ~800 kV/cm, enabling a $W_{rec}$ of ~ 6.63 J cm$^{-3}$ with an excellent $\eta$ of ~ 96% in multilayer ceramic capacitors [27]. Gao *et al*. systematically investigated A-site high-entropy $BaTiO_3$ ceramics and achieved a $W_{rec}$ of 5.16 J cm$^{-3}$ with a $\eta$ of 88%, accompanied by ultrafast discharge characteristics ($t_{0.9}$ ~ 68 ns), highlighting the role of entropy-induced chemical disorder in promoting relaxor behavior [28]. More recently, Kong *et al*. demonstrated a high-entropy $BaTiO_3$-based relaxor ceramic exhibiting a remarkably high $W_{rec}$ of 10.9 J cm$^{-3}$ and $\eta$ of 93% at 720 kV cm$^{-1}$, together with excellent temperature stability from −50 to 260 ºC [29]. In $BiFeO_3$-based systems, entropy engineering has also been explored to enhance dielectric and energy storage properties. Li *et al*. developed high-entropy $(1-x)BiFeO_3$–$x(Ba_{0.2}Sr_{0.2}Ca_{0.2}Bi_{0.2}Na_{0.2})TiO_3$ relaxor ferroelectric ceramics and demonstrated that structural modulation enhances relaxation behavior and polarization response. The optimized composition achieved a high $W_{rec}$ of 12.1 J cm$^{-3}$ with an $\eta$ of 86.1% in multilayer ceramic capacitors [30]. Chen *et al*. introduced a B-site multicomponent high-entropy perovskite into BF–BT ceramics, achieving a $W_{rec}$ of 2.4 J cm$^{-3}$ with an $\eta$ of 75% at 190 kV cm$^{-1}$, while significantly enhancing relaxor behavior and mechanical robustness [31]. Li *et al*. reported that introducing high-entropy engineering into relaxor ferroelectric ceramics enables simultaneous enhancement of polarization difference and breakdown strength, thereby delivering exceptional energy-storage performance [32]. In addition, Liu *et al*. reported a high-entropy perovskite modification in $BaTiO_3$–Bi-based ceramics, where the entropy-induced transition from a tetragonal to a pseudo-cubic structure promoted polar nanoregions and enhanced functional responses [33]. Yu *et al*. demonstrated that constructing a polymorphic heterogeneous shell in core–shell dielectrics simultaneously improves recoverable energy density and efficiency, achieving an impressive energy storage performance of 12.7 J cm$^{-3}$ with 87.2% efficiency [34]. Duan *et al*. utilized entropy engineering to induce local polymorphic distortion and heterogeneous polarization configurations, promoting superparaelectric behavior and obtaining an exceptional recoverable energy density of 15.48 J cm$^{-3}$ with a remarkable efficiency of 90.02% [35]. Chen *et al*. introduced a high-entropy strategy to create local polymorphic distortion with multiphase nanoclusters and random oxygen octahedral tilting, achieving a substantial recoverable energy density of 10.06 J $cm^{-3}$ and an ultrahigh efficiency of 90.8% in lead-free relaxor ferroelectrics [36]. Collectively, these studies demonstrate that high configurational entropy can effectively suppress $P_r$, enhance BDS, and stabilize relaxor ferroelectric behavior. Furthermore, studies on high-entropy ceramics and alloys have revealed that complex thermodynamic and kinetic processes, including diffusion retardation, phase stability, and microstructural evolution under extreme conditions, play critical roles in determining their functional properties [37][38]. Nevertheless, achieving simultaneously ultrahigh BDS and polymer-comparable $W_{rec}$ in lead-free bulk BF–BT-based ceramics remains a formidable challenge, primarily due to the intrinsically strong ferroelectricity and limited electrical robustness of this system.

In this work, we introduce a multicomponent high-entropy perovskite oxide, $Ba(Ti_{0.2}Zr_{0.2}Sn_{0.2}Hf_{0.2}Nb_{0.1}Sc_{0.1})O_3$ (BTHE), into the $0.7BiFeO_3$–$0.3BaTiO_3$ (BF–BT) matrix to construct BF–BT–xBTHE high-entropy ferroelectric ceramics. By rationally exploiting configurational entropy effects, lattice distortion, and microstructural engineering, a synergistic enhancement of relaxor behavior and BDS is achieved. The underlying structure–property–performance relationships are systematically investigated, and the role of entropy engineering in suppressing dielectric breakdown is further elucidated through phase-field simulations.

## 2. Experimental

$BaCO_3$, $Bi_2O_3$, $TiO_2$, $Fe_2O_3$, $HfO_2$, $ZrO_2$, $SnO_2$, $Nb_2O_5$, and $Sc_2O_3$ were used as starting materials. Starting materials were weighed according to the designed stoichiometric ratios ($(0.7BiFeO_3–0.3BaTiO_3)–xBa(Ti_{0.2}Zr_{0.2}Sn_{0.2}Hf_{0.2}Nb_{0.1}Sc_{0.1})O_3$, x= 0, 0.1, 0.3, 0.5) and dispersed in ethanol and subjected to wet ball milling with zirconia milling media for 24 h to achieve uniform mixing. After drying at 80 ºC, we calcined the mixed powders at 750 ºC for 2 hours in a sealed alumina crucible to promote the formation of the perovskite phase. The calcined powders were subsequently subjected to a second ball-milling process for 12 hours to further homogenize the particle distribution. A polyvinyl butyral (PVB) binder with a content of 10 wt.% was added, and the resulting powders were uniaxially pressed into disk-shaped green compacts with a diameter of 10 mm under a pressure of 127 MPa for 20 s. Binder removal was performed by heating the green pellets at 600 ºC for 2 hours, during which the samples were placed on a zirconia setter and covered with alumina crucibles to suppress volatilization. Final sintering was performed at temperatures ranging from 1000 to 1100 ºC for 3 hours to obtain dense ceramic bodies. For electrical measurements, silver paste was applied to both surfaces of the sintered samples and fired at 850 ºC for 30 mins to form electrodes. For unipolar ferroelectric measurements, the ceramic samples were mechanically ground and polished to a thickness below 50 μm, followed by the deposition of gold electrodes on both surfaces. The gold electrodes for unipolar *P–E* measurements were deposited on polished samples with an effective electrode diameter of approximately 1 mm, while the silver electrodes for dielectric and breakdown measurements covered the entire sample surface.

Phase structures were characterized by X-ray diffraction (XRD, Empyrean, Panalytical, Netherlands), and Rietveld refinement was performed to quantify phase fractions and lattice parameters. XRD patterns were collected using Cu $K\alpha$ radiation over a $2\theta$ range of 20–80° at a scanning rate of 5° $min^{-1}$. Rietveld refinements were conducted using GSAS II software. Microstructural features were examined using scanning electron microscopy (SEM, Carl Zeiss, Germany), while elemental distributions were analyzed by energy-dispersive X-ray spectroscopy (EDS, Carl Zeiss, Germany). Impedance spectroscopy and dielectric properties (E4980A, Agilent Technologies, USA) were measured over a wide temperature and frequency range (DMS-2000, Balab Technologies, China). Ferroelectric and energy storage behaviors were evaluated using bipolar and unipolar polarization–electric field (*P*–*E*) measurements (PK-10E, PolyK Technologies, USA). BDS was statistically analyzed using Weibull distribution. Phase-field simulations were employed to qualitatively analyze dielectric breakdown evolution.

## 3. Results and Discussion

Figure 1a presents the XRD patterns of BF–BT and BF–BT–xBTHE (x = 0.1, 0.3, and 0.5) ceramics, together with enlarged views of selected diffraction peaks. According to the actual chemical formula, the average B-site configurational entropy increases from 1.14R (x = 0.1) to 1.69R (x = 0.5), while the x = 0.3 composition reaches the typical high-entropy criterion ($\Delta S/R \approx 1.5$). All compositions exhibit a single perovskite phase without detectable secondary phases, indicating that the multicomponent BTHE species are successfully incorporated into the BF–BT lattice to form homogeneous solid solutions. With increasing x, the diffraction peaks gradually shift toward lower angles, suggesting a progressive lattice expansion induced by high-entropy substitution at the B site.

The evolution of the Bragg (200) reflection further reveals a composition-dependent structural change. The undoped BF–BT matrix exhibits a relatively broad and quasi-symmetric profile, characteristic of a pseudo-cubic structure near the morphotropic phase boundary. Upon incorporation of the BTHE components, a discernible peak splitting gradually appears, reflecting the enhanced lattice distortion and local strain fluctuations introduced by chemical disorder.

To quantitatively analyze the structural evolution, Rietveld refinements were performed for all compositions, as shown in Figures 1b–e, and the corresponding refinement parameters are summarized in Table 1. The BF–BT sample is well described by a coexisting rhombohedral (R) and cubic (C) phase model. With increasing x, the fraction of the R phase decreases markedly, and for x = 0.3 and 0.5, the diffraction patterns can be refined using a single cubic phase, indicating the stabilization of an entropy-stabilized C structure. It should be noted that the observed increase of the rhombohedral phase fraction at low BTHE content is associated with enhanced local lattice distortion and phase competition near the morphotropic phase boundary. As configurational entropy increases further, long-range ferroelectric ordering is progressively suppressed, stabilizing the pseudo-cubic structure.

Consistent with the peak shifts observed in Figure 1a, the refined lattice parameters and unit cell volumes in Table 1 show a monotonic increase with increasing BTHE content. Both the R and C unit-cell volumes expand progressively, which can be attributed to the incorporation of relatively larger B-site cations such as $Zr^{4+}$, $Hf^{4+}$, $Sn^{4+}$, and $Sc^{3+}$ compared to the host cations $Ti^{4+}$ and $Fe^{3+}$. This lattice expansion, together with the reduced phase asymmetry, confirms that high-entropy engineering introduces significant chemical disorder and local structural fluctuations, thereby establishing a structurally disordered framework for the subsequent dielectric and energy-storage behaviors. Although the $R_{wp}$ values are relatively high, this behavior is common in high-entropy systems characterized by severe lattice distortion and broadened diffraction peaks. The absence of impurity peaks and the satisfactory agreement between observed and calculated profiles indicate that the refinements reliably evaluate the structural evolution of the present ceramics.

SEM images reveal a systematic and pronounced grain refinement trend with increasing BTHE incorporation, as shown in Figure 2. The pristine BF–BT ceramic is characterized by coarse grains in the range of 10–20 μm, whereas the introduction of the high-entropy perovskite component leads to a continuous reduction in grain size. Specifically, the average grain size decreases markedly upon low-level BTHE addition and is further suppressed at intermediate and high entropy contents, ultimately reaching only a few micrometers in the BF–BT–0.5BTHE composition. Consistent with this trend, the corresponding grain size distribution histograms (Figure 2) exhibit a progressive narrowing and a distinct shift toward finer grains, indicating a substantial increase in grain-boundary density in the high-entropy ceramics.

Elemental homogeneity was further examined using EDS elemental mapping for the representative BF–BT–0.3BTHE sample (Figure 3). All constituent elements, including Bi, Ba, Ti, Fe, Sc, Nb, Zr, Sn, and Hf, are uniformly distributed throughout the ceramic matrix without any detectable elemental segregation or secondary-phase enrichment, confirming the formation of a chemically homogeneous high-entropy perovskite structure.

The observed grain refinement is closely associated with the sluggish diffusion effect induced by high configurational entropy, which effectively retards atomic diffusion and grain coarsening during sintering. Such a microstructural evolution is highly beneficial for dielectric energy-storage applications, as the increased grain-boundary density effectively impedes charge-carrier migration and disrupts the development of continuous electrical breakdown pathways. As a result, the refined microstructure provides a critical structural basis for the significantly enhanced BDS observed in the BF–BT–xBTHE ceramics.

The temperature-dependent dielectric spectra of BF–BT–xBTHE ceramics exhibit a significantly broadened dielectric permittivity peak and pronounced frequency dispersion, which are characteristic features of relaxor ferroelectrics, as shown in Figure 4 [39]. A quantitative analysis based on the modified Curie–Weiss law, expressed as

$$\frac{1}{\varepsilon}-\frac{1}{\varepsilon_{\mathrm{max}}}=\frac{(T-T_{\mathrm{m}})^{\gamma}}{C} \quad (2)$$

where $\varepsilon_{max}$ is the maximum dielectric permittivity at the temperature $T_{max}$, $C$ is the Curie-like constant, and $\gamma$ denotes the diffuseness parameter [40], reveals $\gamma$ values exceeding 1.5 for the high-entropy compositions (Figure 5), confirming the occurrence of a diffuse phase transition. This relaxation behavior originates from compositional disorder and local structural heterogeneity induced by multicomponent B-site substitution, which disrupts long-range ferroelectric ordering and promotes nanoscale polarization fluctuations.

Impedance spectroscopy was further employed to elucidate the electrical transport and relaxation characteristics of BF–BT–xBTHE ceramics. As shown in Figure 6b, the frequency-dependent curves of the imaginary part of impedance (Z'') and the electric modulus (M'') reveal a single and well-defined relaxation peak, whose positions largely overlap within the investigated frequency range. The high consistency between the Z'' and M'' responses indicates highly uniform electrical behavior [41], which is primarily governed by a single relaxation process rather than by multiple competing conduction or interfacial polarization mechanisms. The impedance spectra exhibit a single depressed semicircle in the Nyquist plots, indicating that the electrical response is dominated by one relaxation process (Figure 6a). The depressed nature of the semicircle suggests non-Debye relaxation behavior arising from entropy-induced lattice distortion and local structural heterogeneity. Consequently, the impedance response can be effectively modeled by an equivalent circuit comprising a bulk resistance ($R_b$) in parallel with a constant phase element (CPE). This electrical homogeneity effectively suppresses local electric field distortion and parasitic loss paths, which is crucial for achieving both high BDS and high $\eta$ of energy storage in dielectric ceramics. The activation energy obtained from Arrhenius analysis is approximately 1.05–1.16 eV (Figure 6c and 6d), suggesting that the conduction process likely involves oxygen-vacancy-related migration, consistent with reported activation energy ranges in similar perovskite systems [42][43]. With increasing BTHE content, the bulk resistance gradually increases, indicating that charge-carrier mobility is significantly suppressed. This can be attributed to lattice potential distortion and defect pinning effects caused by entropy engineering. The synergistic modulation of dielectric relaxation and electrical transport provides an important microscopic basis for the enhanced dielectric reliability and superior energy storage performance observed in high-entropy BF-BT ceramics. It is important to emphasize that enhanced energy storage performance arises not from a single “entropy effect" but from multiple synergistic mechanisms. Specifically, high configurational entropy induces lattice distortion, sluggish diffusion, and defect modulation, collectively contributing to grain refinement, suppression of electrical conduction, and homogenization of local electric fields. These factors ultimately lead to improved breakdown strength and energy storage performance.

The bipolar *P–E* hysteresis loops of BF–BT–xBTHE ceramics gradually evolve from typical ferroelectric behavior with large hysteresis in the undoped sample to slim, elongated loops with increasing BTHE content (Figure 7a). Correspondingly, unipolar P–E loops exhibit a substantial reduction in $P_r$ while maintaining high maximum polarization ($P_{max}$), which is highly desirable for energy storage applications (Figure 7b). Benefiting from the synergistic optimization of relaxor behavior and BDS, the BF–BT–0.3BTHE composition achieves an ultrahigh BDS of 840 kV $cm^{-1}$. Under this electric field, a remarkable $W_{rec}$ of 10.55 J $cm^{-3}$ is obtained, accompanied by a high energy $\eta$ of 79%. Notably, this performance represents a significant advancement for bulk lead-free ceramic dielectrics. Compared to previously reported bulk systems, which often exhibit limited breakdown strength, the present work achieves a superior combination of high BDS and high $W_{rec}$ (Figure 8). Notably, many reported ultrahigh energy densities are achieved in multilayer capacitors or thin films, whereas attaining comparable performance in bulk ceramics remains considerably more challenging.

The BDS of the BF–BT–xBTHE ceramics was statistically analyzed using the Weibull distribution. For each composition, the experimental breakdown data were ranked in ascending order, and the Weibull coordinates are calculated according to

$$X_i = \ln\left(E_i\right) \quad (3)$$

$$Y_i = \ln\left[-ln\left(1-\frac{i}{n+1}\right)\right] \quad (4)$$

where $E_i$ is the BDS of the $i$-th specimen and $n$ denotes the total number of tested samples [44]. A linear relationship between $Y_i$ and $X_i$ indicates that the breakdown behavior follows Weibull statistics, with the slope corresponding to the Weibull modulus and the intercept related to the characteristic BDS. The high-entropy ceramics exhibit comparable Weibull slopes with moderate data scatter, indicating a stable and reproducible breakdown behavior. The characteristic BDS increases with BTHE incorporation within an appropriate entropy range, which can be associated with the combined effects of grain refinement, reduced electrical conduction, and a more uniform electric field distribution. As shown in Figure 9, the Weibull distribution of BDS for BF–BT–xBTHE (x = 0.1, 0.3, 0.5) illustrates this trend. To further elucidate the role of entropy engineering, the configurational entropy and key physical parameters are summarized in Table 2.

COMSOL Multiphysics and MATLAB simulations were employed to qualitatively elucidate the breakdown suppression mechanism induced by entropy engineering [45]. As shown in Figure 10a and 10b, the introduction of high-entropy-induced microstructural heterogeneity results in a significantly more homogeneous local electric field distribution compared with the pristine BF–BT system. The formation of distributed electric field regions effectively mitigates local field concentration, which is widely recognized as the initiation site of dielectric breakdown. Moreover, the simulated breakdown evolution reveals highly tortuous and branched breakdown pathways in the high-entropy ceramic (Figure 10c and 10d), in contrast to the relatively straight and localized breakdown channel in the undoped counterpart. Such dispersed breakdown behavior significantly delays the formation of a continuous conductive path, providing a physical explanation for the experimentally observed enhancement in BDS.

The structure–property–performance relationship can be summarized as follows: entropy-induced lattice distortion and pseudo-cubic phase formation promote relaxor behavior, while grain refinement and chemical homogeneity suppress electrical conduction. Collectively, these effects lead to a more uniform electric field distribution and delayed breakdown, ultimately enhancing energy storage performance.

## 4. Conclusion

In conclusion, we successfully employed an entropy-engineering strategy to develop BF–BT–xBTHE lead-free high-entropy ferroelectric ceramics with outstanding energy storage performance. The introduction of multicomponent B-site cations induces severe lattice distortion, relaxor ferroelectric behavior, and significant grain refinement, collectively leading to suppressed electrical conduction and enhanced BDS. An ultrahigh $W_{rec}$ of 10.55 J cm$^{-3}$ with a high $\eta$ of 79% is achieved at a BDS of 840 kV cm$^{-1}$, representing a breakthrough for bulk lead-free ceramic dielectrics.

CRediT authorship contribution statement

**Yitao Jiao:** Writing – original draft, Investigation, Project administration, Methodology, Validation, Data curation. **Zhenhao Fan:** Investigation. **Hai-Feng Li:** Writing – review & editing, Supervision. **Dawei Wang:** Writing – review & editing, Supervision, Resources, Funding acquisition.

## Conflicts of interest

The authors declare that they have no known competing financial interests or personal relationships that could have appeared to influence the work reported in this paper.

## Acknowledgements

This research is supported by the Natural Science Foundation of China (No. 52472122).

## Figure and table list

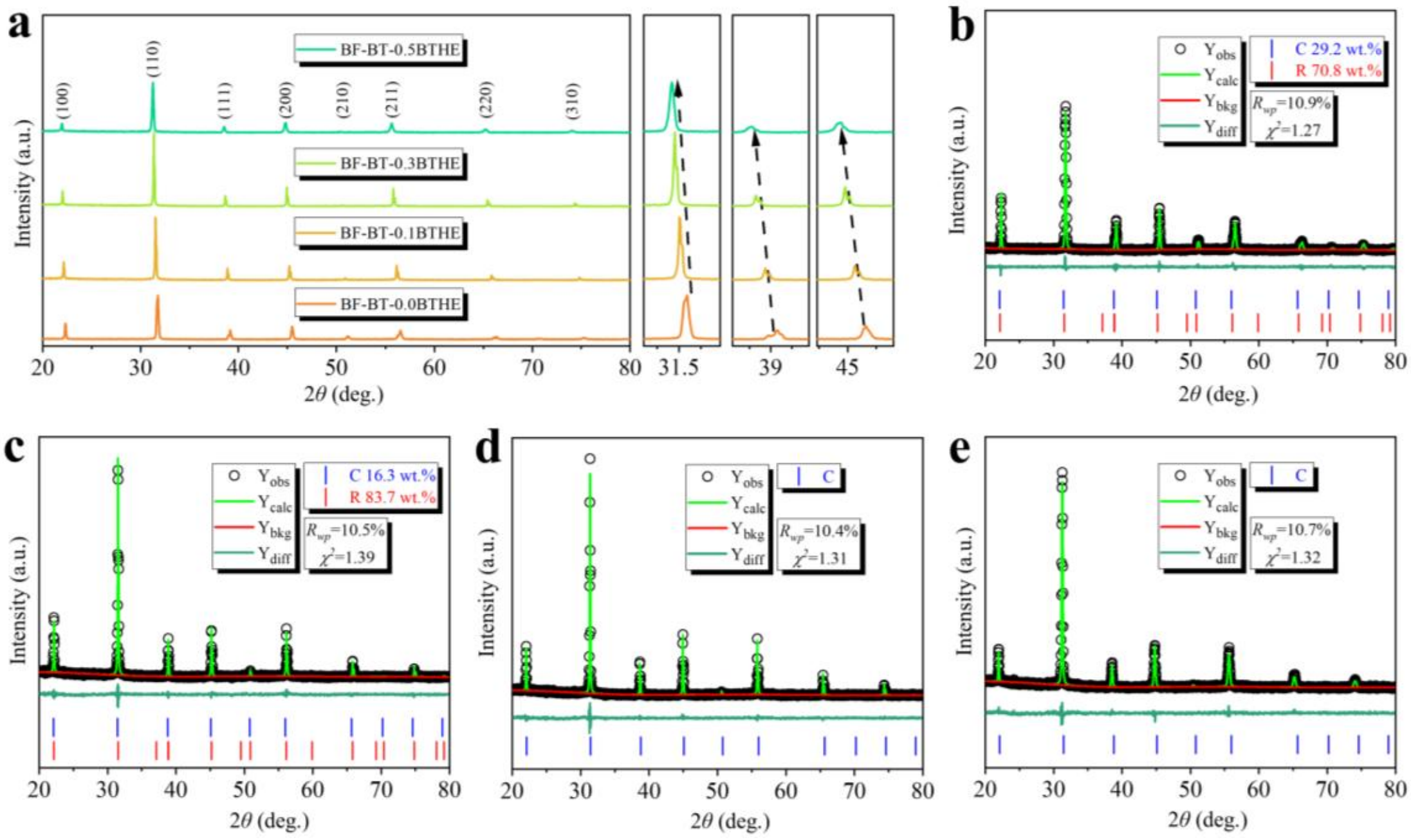

Fig. 1. (a) XRD patterns of BF–BT and BF–BT–xBTHE (x= 0.1, 0.3, 0.5) ceramics with enlarged XRD profiles. (b–e) Rietveld refinement results of BF–BT (b) and BF–BT–xBTHE ceramics with x = 0.1 (c), 0.3 (d), and 0.5 (e). (C: Cubic; R: Rhombohedral)

Table 1. The Rietveld refinement results for BF–BT-xBTHE ceramics

| Phase fraction (wt%) | | | Rhombohedral (R) | | | Cubic (C) | |
|---|---|---|---|---|---|---|---|
| x | R | C | *a*=*b* (Å) | *c* (Å) | *V* (Å$^3$) | *a*=*b*=*c* (Å) | *V* (Å$^3$) |
| 0 | 70.8 | 29.2 | 5.6286(0) | 13.8715(1) | 380.592(0) | 3.9879(0) | 63.422(0) |
| 0.1 | 83.7 | 16.3 | 5.6687(0) | 13.8850(1) | 386.412(0) | 4.0174(0) | 64.840(0) |
| 0.3 | | 100 | | | | 4.0315(1) | 65.528(0) |
| 0.5 | | 100 | | | | 4.0497(1) | 66.418(0) |

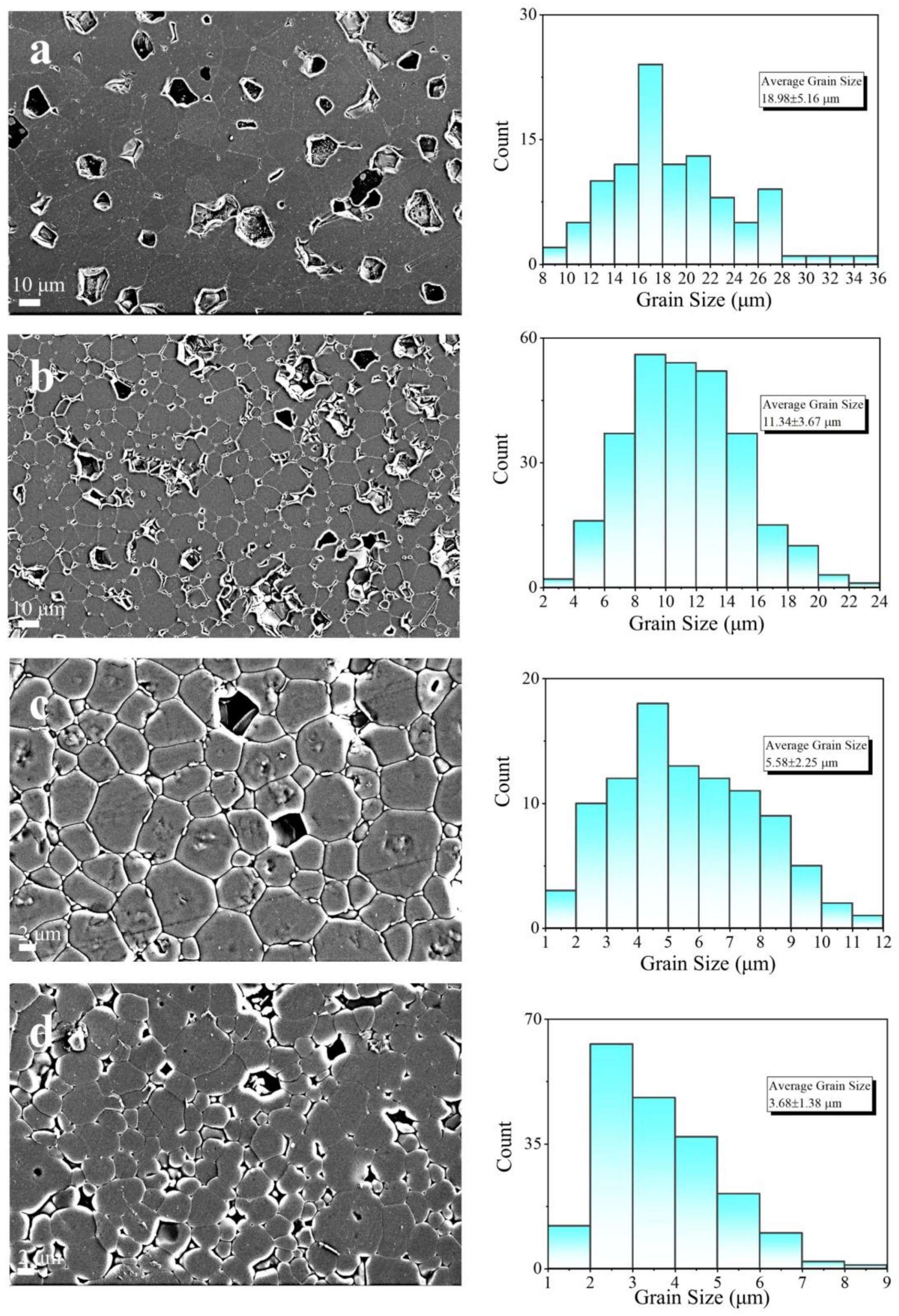

Fig. 2. SEM images of (a) BF–BT, (b) BF–BT–0.1BTHE, (c) BF–BT–0.3BTHE, and (d) BF–BT–0.5BTHE ceramics, showing progressive grain refinement with increasing BTHE content.

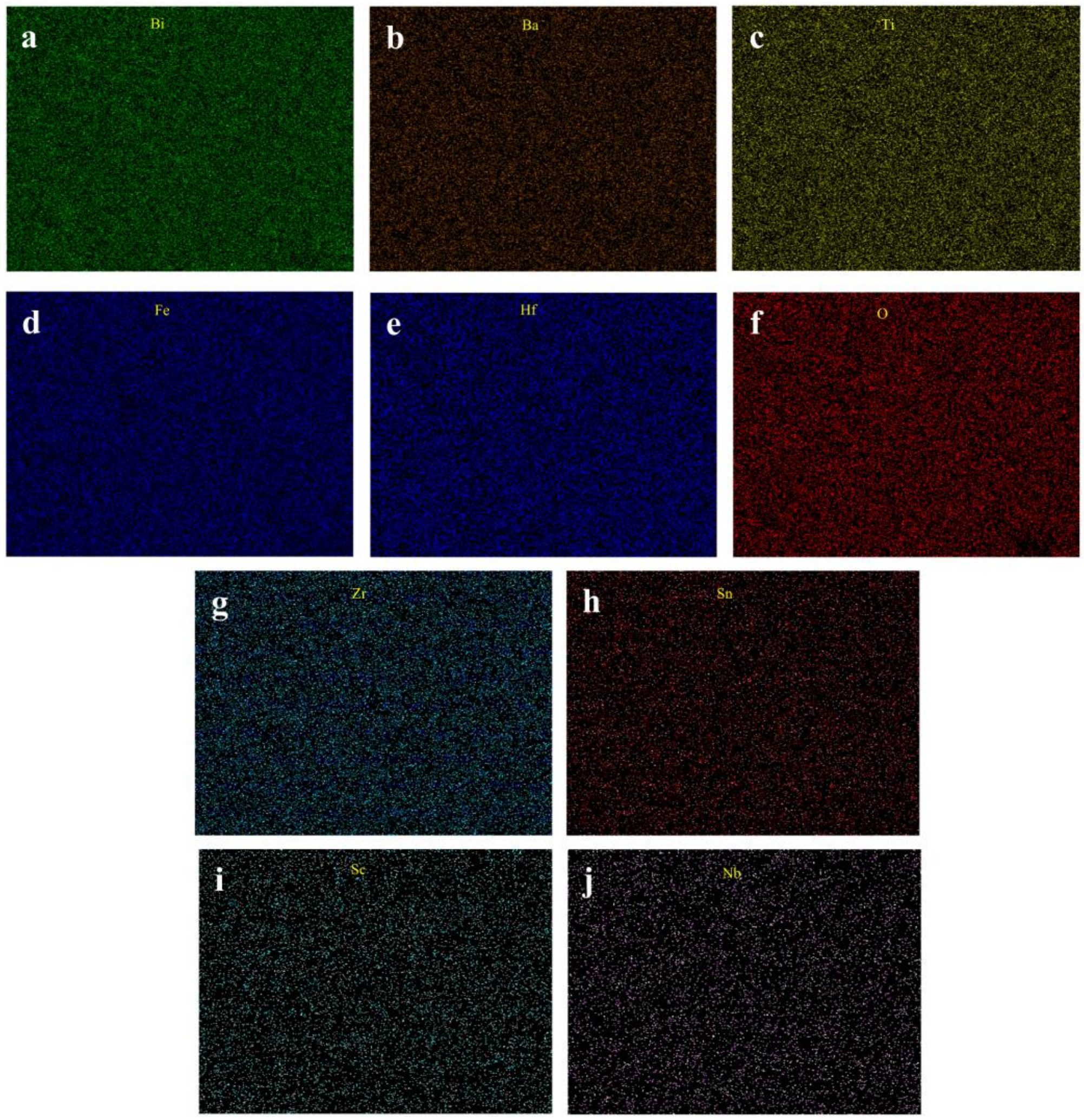

Fig. 3. EDS elemental mapping of the representative BF–BT–0.3BTHE ceramic, demonstrating a homogeneous spatial distribution of Bi, Ba, Ti, Fe, Sc, Nb, Zr, Sn, and Hf without detectable elemental segregation.

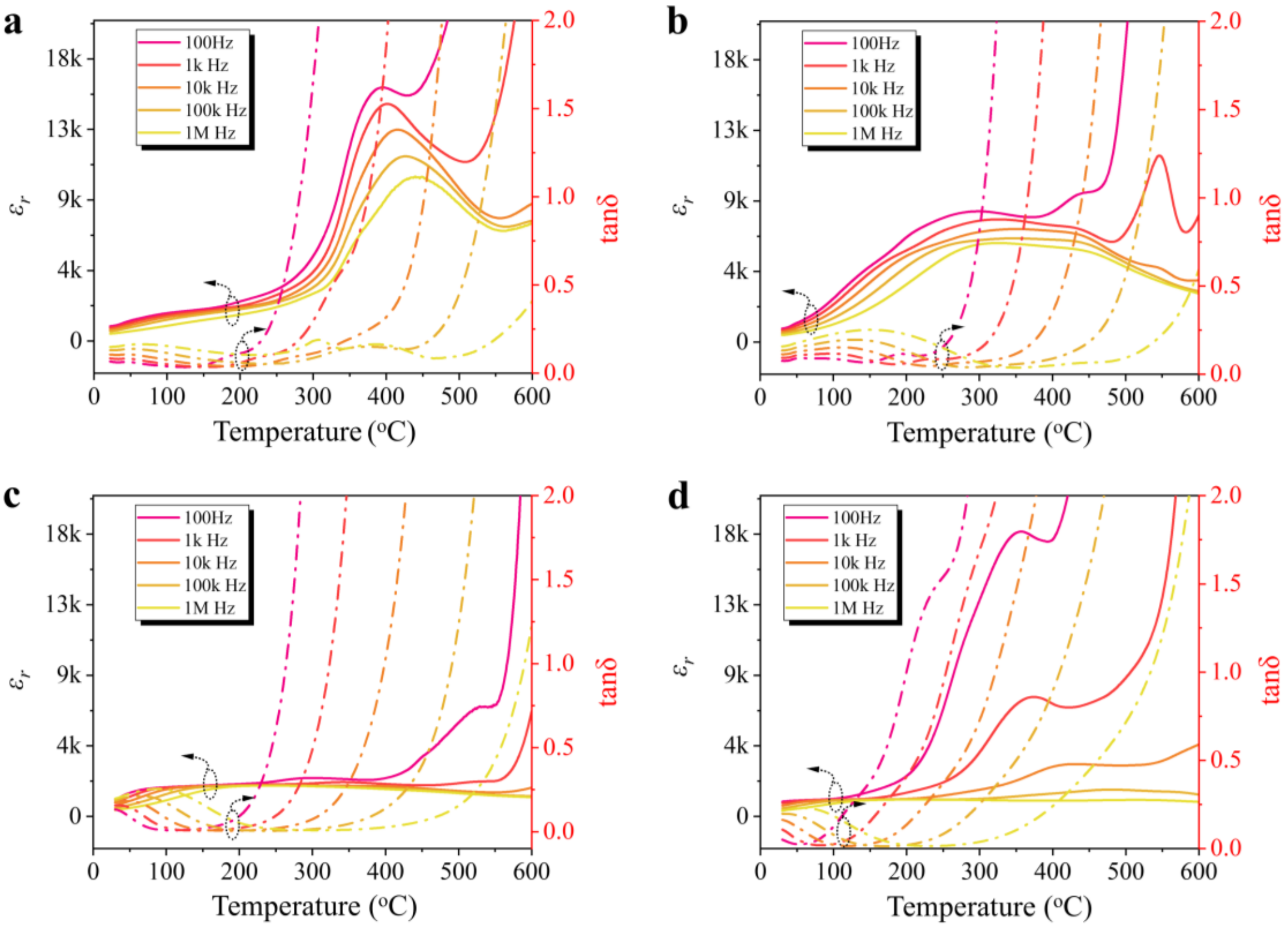

Fig. 4. Temperature and frequency dependence of the dielectric constant ($\varepsilon_r$) and dielectric loss (tanδ) for the BF-BT (a), BF-BT-0.1BTHE (b), BF-BT-0.3BTHE (c) and BF-BT-0.5BTHE (d) ceramics.

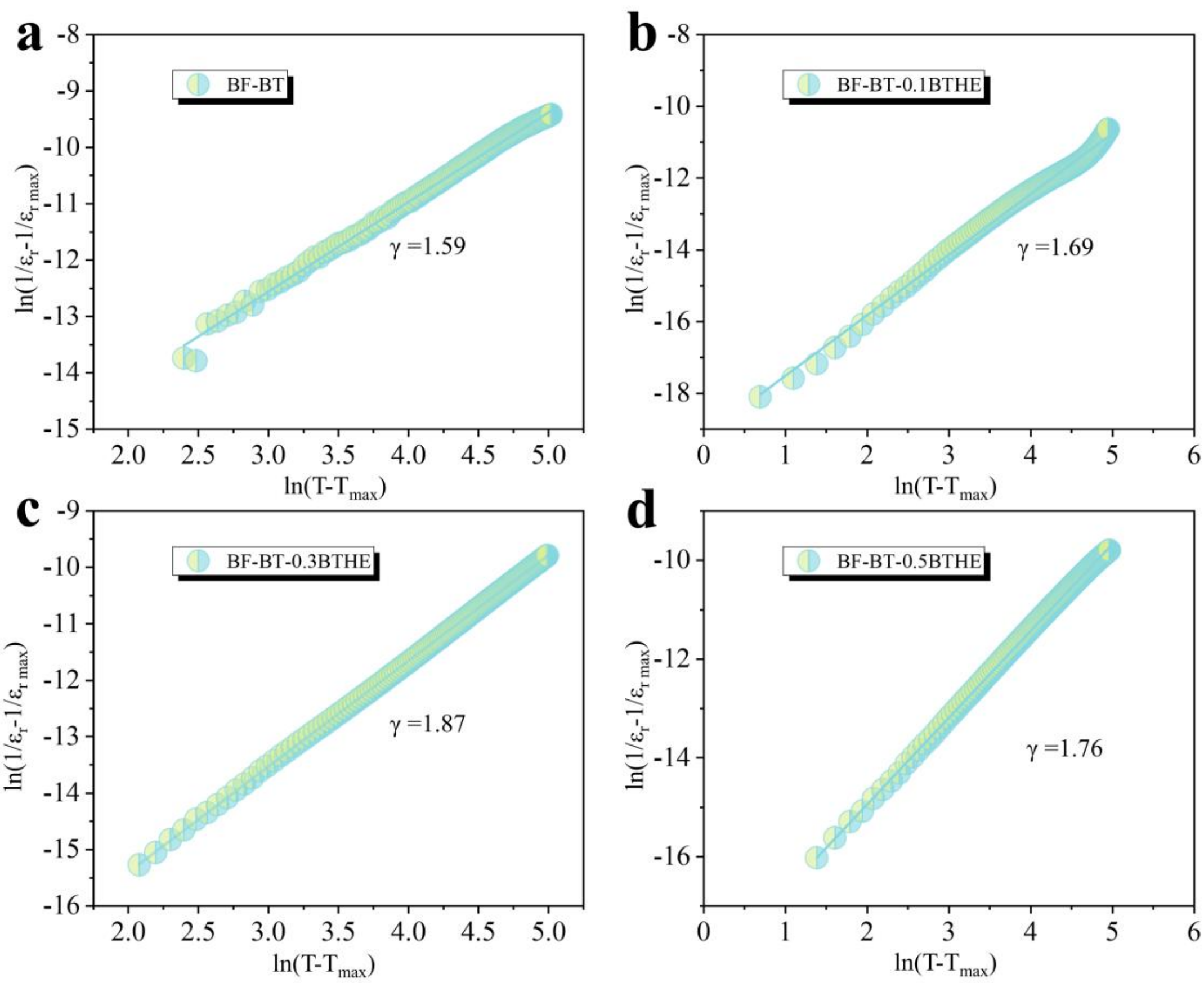

Fig. 5. Quantitative analysis of the relaxor behavior in BF-BT (a), BF-BT-0.1BTHE (b), BF-BT-0.3BTHE (c) and BF-BT-0.5BTHE (d) ceramics.

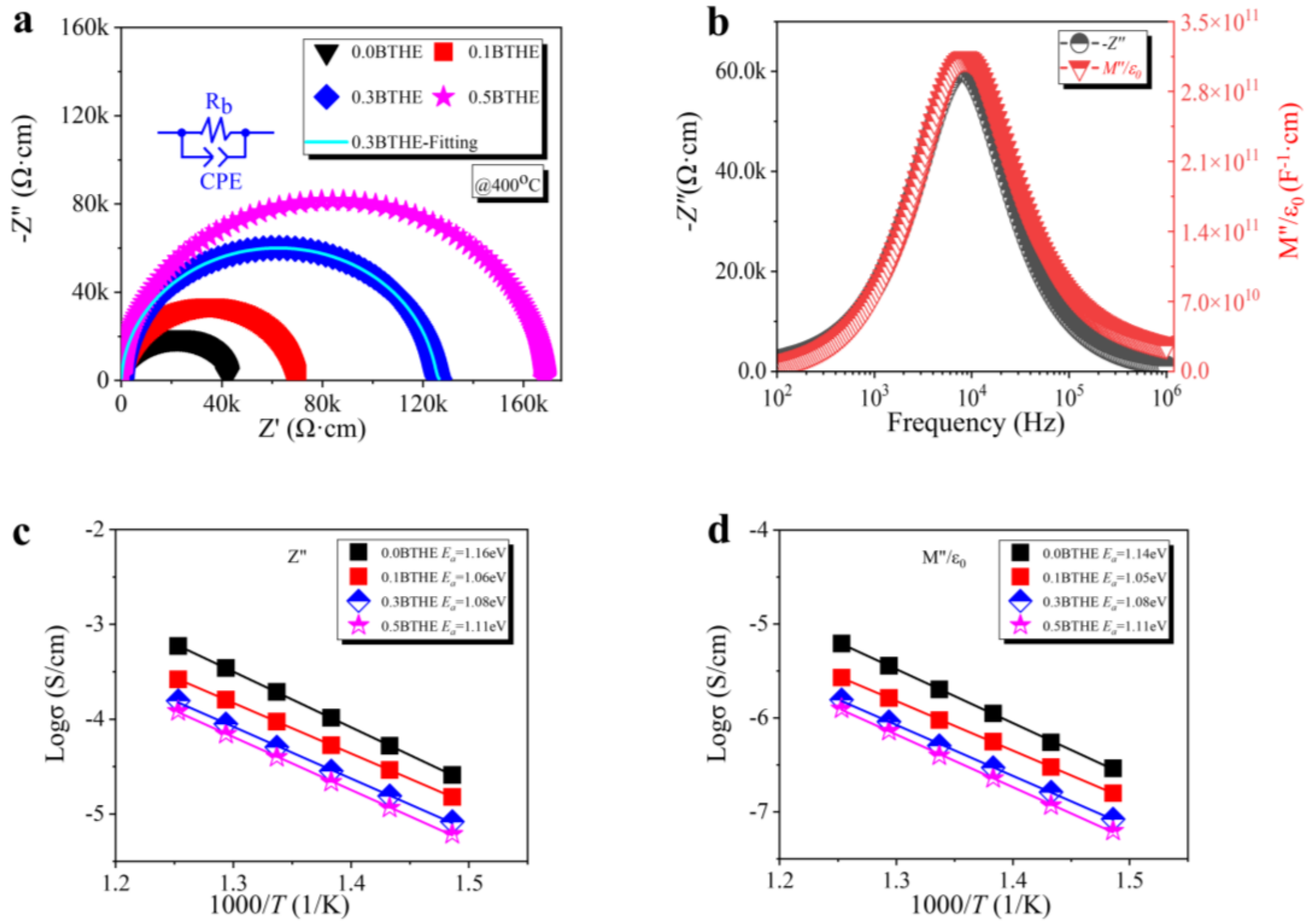

Fig. 6. (a) Experimental complex impedance (Z′–Z″) spectra of BF–BT–xBTHE ceramics measured at elevated temperature and fitted using an equivalent circuit model of $R_b$‖CPE. (b) Frequency dependence of the imaginary parts of impedance (*Z″*) and electric modulus (*M″*) for a representative composition. (c, d) Arrhenius plots derived from (c) impedance and (d) modulus relaxation processes, respectively.

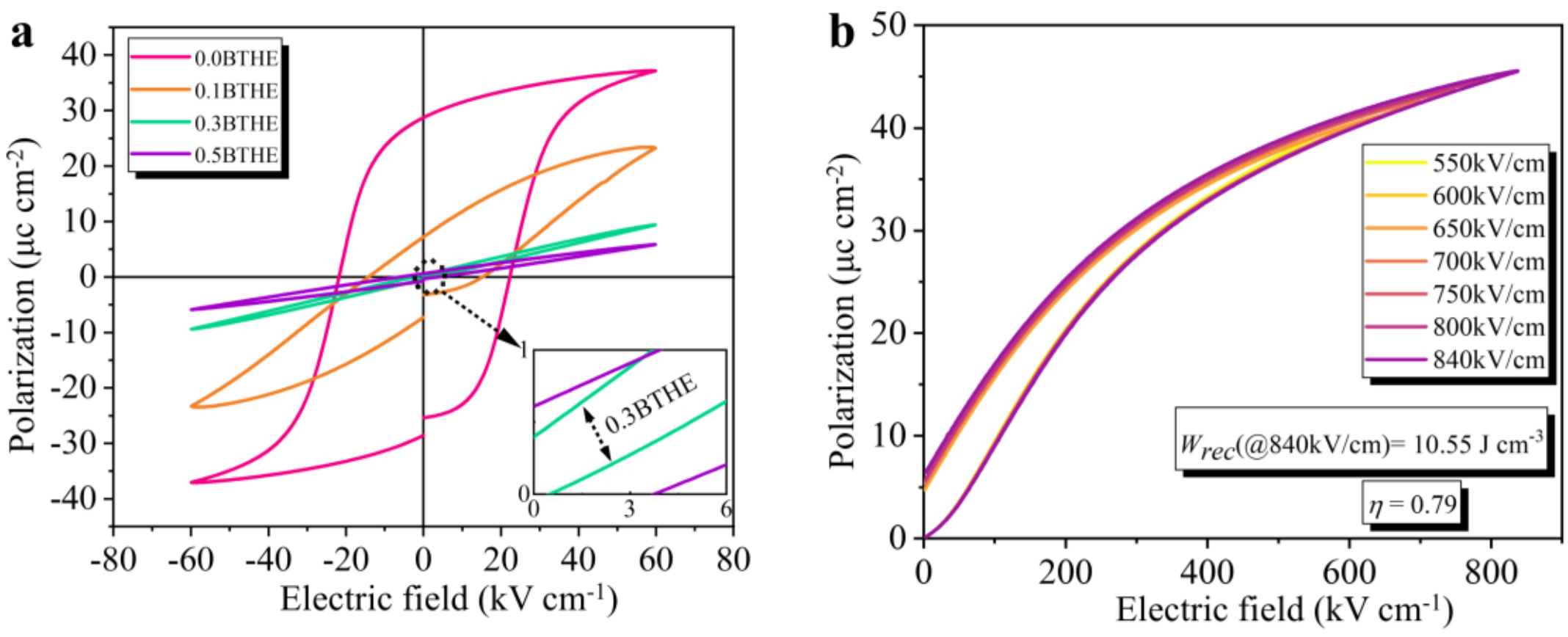

Fig. 7. (a) Bipolar *P*-*E* hysteresis loop. (b) Unipolar *P*–*E* hysteresis loops of BF–BT–0.3BTHE.

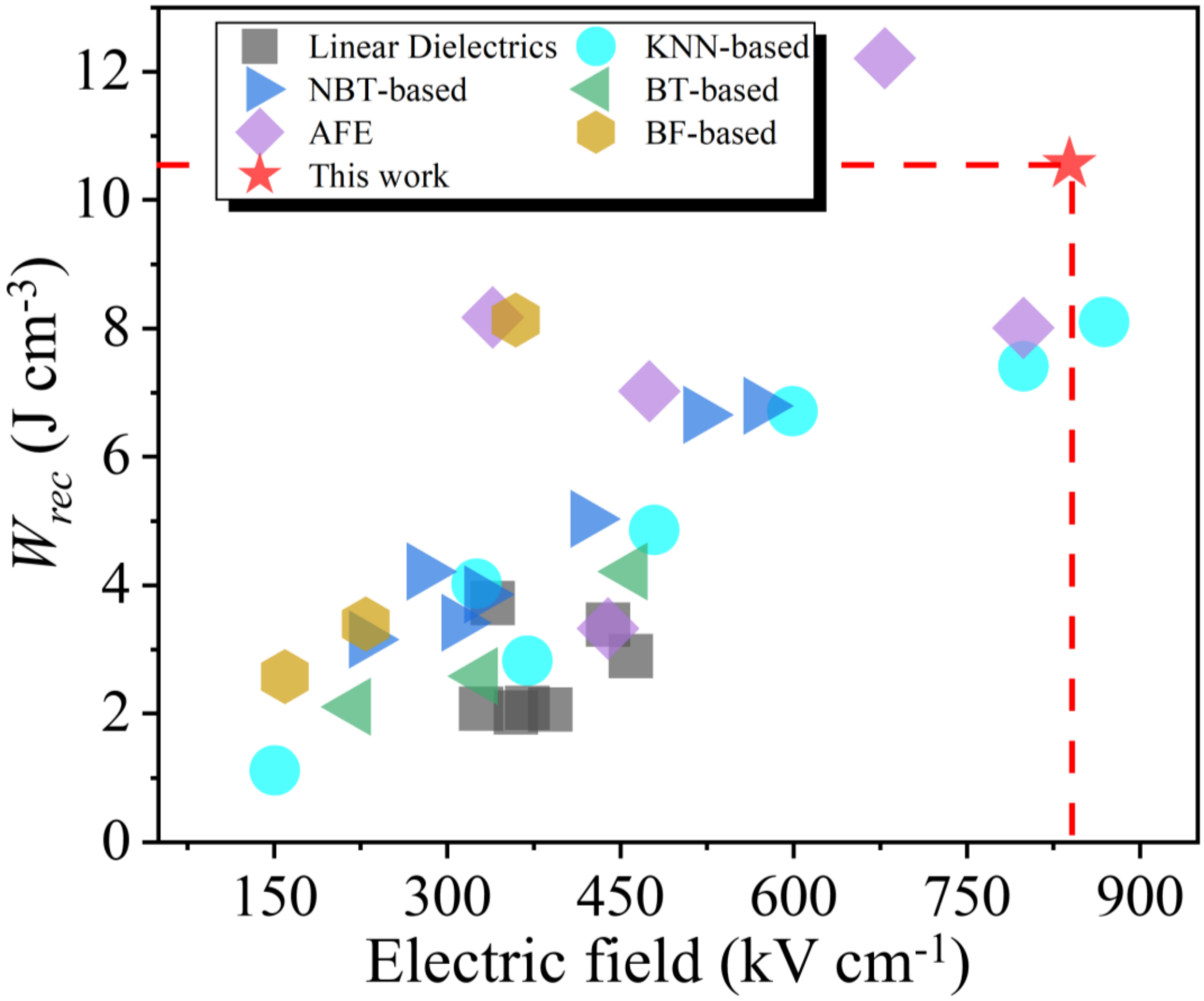

Fig. 8. Comparison of recoverable energy density ($W_{rec}$) and breakdown strength (BDS) between the present BF–BT–0.3BTHE ceramic and representative lead-free bulk dielectric ceramics reported in recent literature [46][47]. (KNN = (K,Na)$NbO_3$; NBT = $Na_{0.5}Bi_{0.5}TiO_3$; BT = $BaTiO_3$; BF = $BiFeO_3$; AFE = antiferroelectric.)

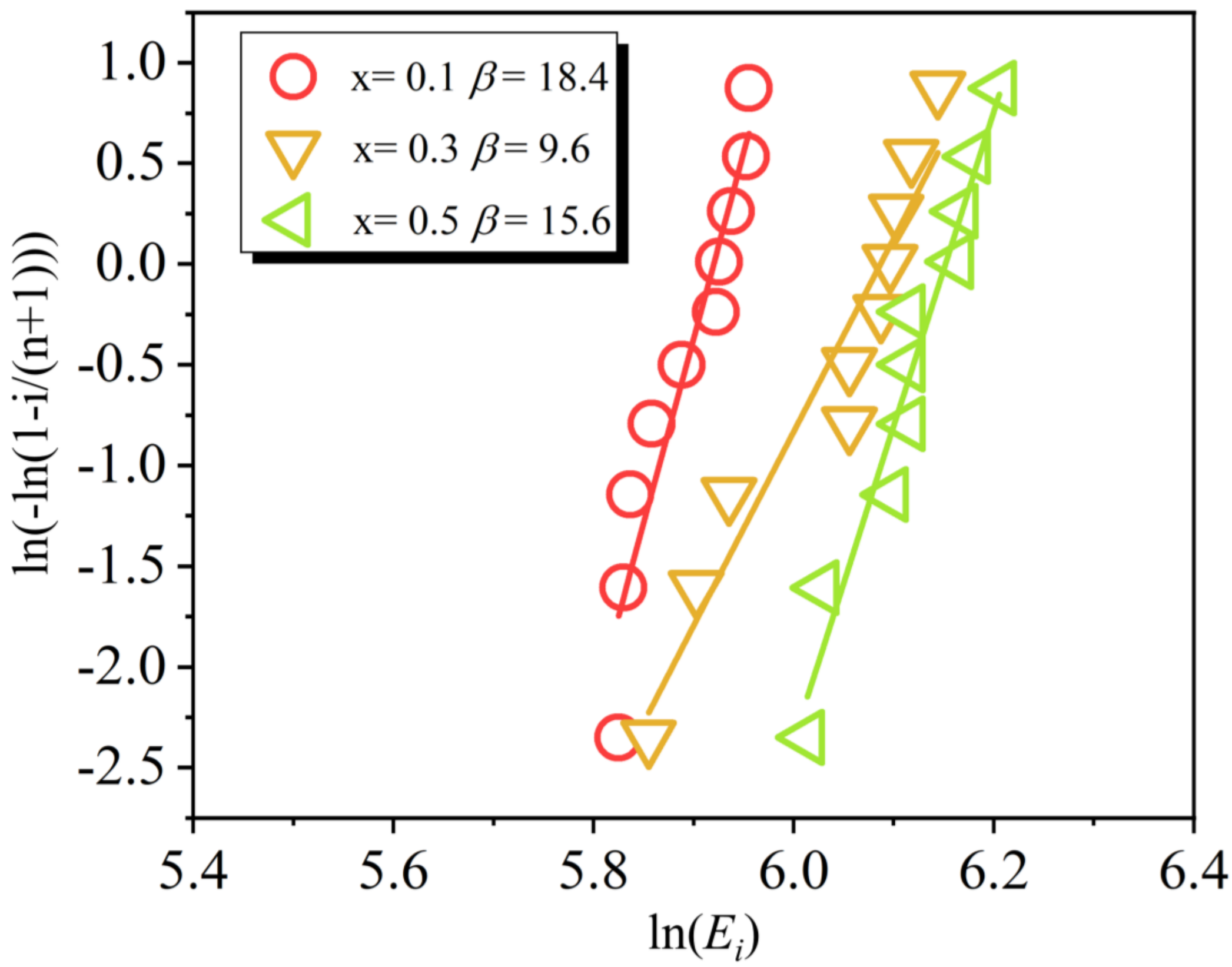

Fig. 9. Weibull distribution of BDS for BF–BT–xBTHE (x= 0.1, 0.3, 0.5).

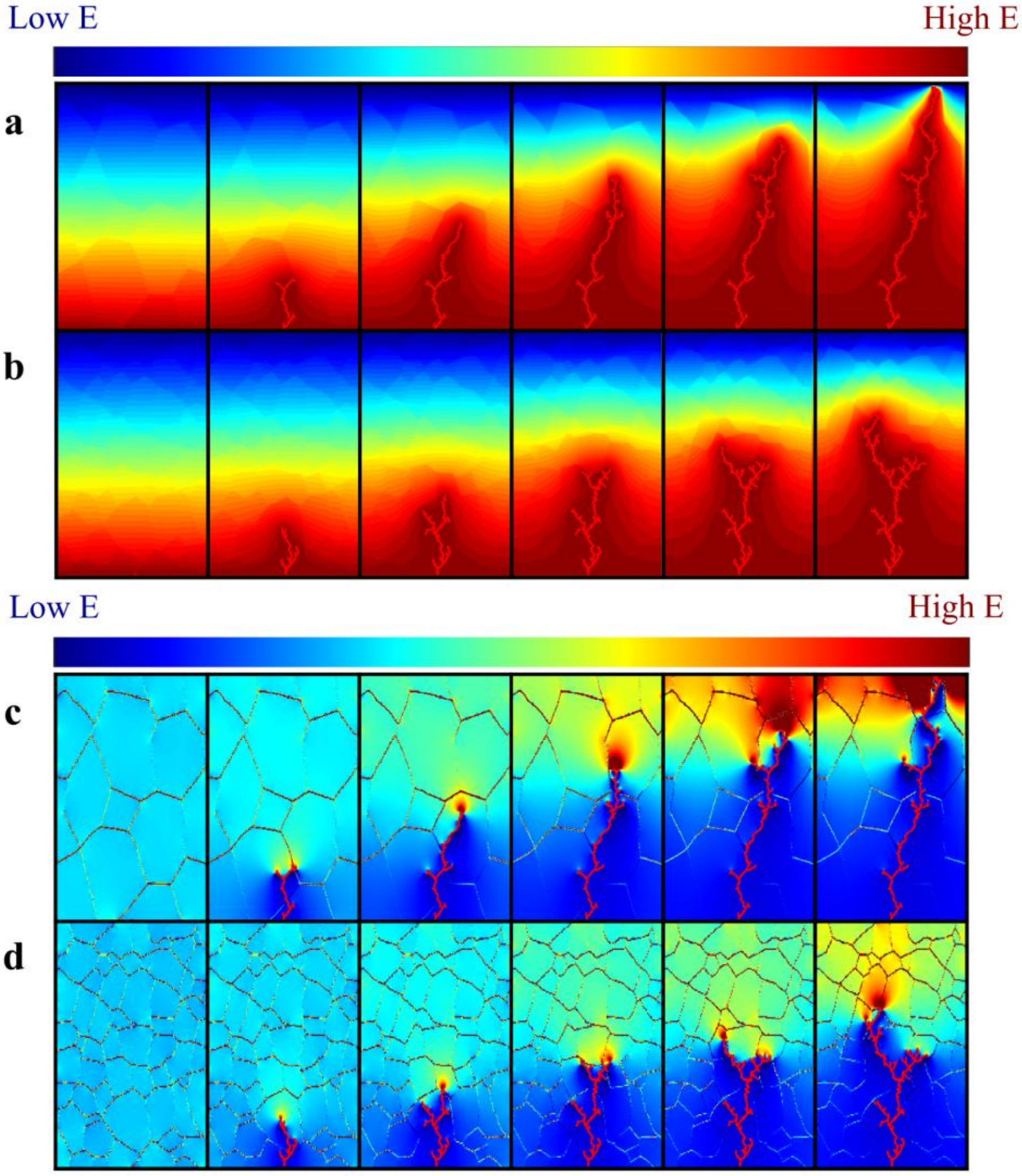

Fig. 10. Electric potential distributions with electrical tree evolution for (a) BF-BT and (b) BF-BT-0.3BTHE ceramics. Electric field distributions with electrical tree evolution for (c) BF-BT and (d) BF-BT-0.3BTHE.

Table 2 Correlation between configurational entropy and characteristic parameters of BF–BT–xBTHE ceramics.

| Composition | Configurational Entropy (ΔS/R) | Average Grain Size (μm) | Activation Energy from Z'' (eV) | Activation Energy from M'' (eV) | Characteristic BDS (kV $cm^{-1}$) |
|---|---|---|---|---|---|
| BF–BT (x=0) | 0 | 18.98±5.16 | 1.16 | 1.14 | 60 |
| BF–BT–0.1BTHE | 1.14 | 11.34±3.67 | 1.06 | 1.05 | 230 |
| BF–BT–0.3BTHE | 1.50 | 5.58±2.25 | 1.08 | 1.08 | 840 |
| BF–BT–0.5BTHE | 1.69 | 3.68±1.38 | 1.11 | 1.11 | 500 |